\documentclass[conference]{IEEEtran}
\IEEEoverridecommandlockouts
\usepackage{cite}
\usepackage{geometry}
\usepackage{amsmath,amssymb,amsfonts}
\usepackage{algorithmic}
\usepackage{graphicx}
\usepackage{textcomp}
\usepackage{xcolor}
\usepackage{stfloats}
\usepackage{algorithm}
\usepackage{caption}
\usepackage{epstopdf}
\usepackage{subfigure}
\usepackage{bm}
\usepackage{amsmath}
\usepackage{algorithmic}
\usepackage{comment}
\usepackage{booktabs}
\usepackage{amsmath}
\usepackage{amsthm}

\allowdisplaybreaks[4]

\newtheoremstyle{customtheorem}
{3pt}
{3pt}
{}
{1em}
{\bfseries}
{:}
{0.5em}
{}

\theoremstyle{customtheorem}

\newtheorem{theorem}{Theorem}

\makeatletter
\renewenvironment{proof}[1][\proofname]
{\par\pushQED{\qed}\normalfont\topsep6\p@\@plus6\p@\relax
	\hspace*{1em} 
	\trivlist\item[\hskip\labelsep\hspace*{1em}\bfseries 
	#1\@addpunct{:}]\ignorespaces}
{\popQED\endtrivlist\@endpefalse}
\makeatother

\usepackage{caption}
\begin{document}

\title{Resource Allocation Based on Optimal Transport Theory in ISAC-Enabled Multi-UAV Networks\\
}

\author{
	\IEEEauthorblockN{
		Yufeng Zheng\IEEEauthorrefmark{1}, 
		Lixin Li\IEEEauthorrefmark{1}, 
		Wensheng Lin\IEEEauthorrefmark{1}, 
		Wei Liang\IEEEauthorrefmark{1}, 
		Qinghe Du\IEEEauthorrefmark{2} 
		and Zhu Han\IEEEauthorrefmark{3}} 
	\IEEEauthorblockA{\IEEEauthorrefmark{1}School of Electronics and Information, Northwestern Polytechnical University, Xi’an, China, 710129}
	\IEEEauthorblockA{\IEEEauthorrefmark{2}School of Information and Communications Engineering, Xi’an Jiaotong University, Xi’an, China, 710049}
	\IEEEauthorblockA{\IEEEauthorrefmark{3}Department of Electrical and Computer Engineering, University of Houston, Houston, TX, 77004} 
	
	\thanks{This paper has been accepted for publication in IEEE Globecom 2024 workshop.}
} 

\maketitle

\begin{abstract}
This paper investigates the resource allocation optimization for cooperative communication with non-cooperative localization in integrated sensing and communications (ISAC)-enabled multi-unmanned aerial vehicle (UAV) cooperative networks. 
Our goal is to maximize the weighted sum of the system's average sum rate and the localization quality of service (QoS) by jointly optimizing cell association, communication power allocation, and sensing power allocation. Since the formulated problem is a mixed-integer nonconvex problem, we propose the alternating iteration algorithm based on optimal transport theory (AIBOT) to solve the optimization problem more effectively. Simulation results demonstrate that the AIBOT can improve the system sum rate by nearly 12\% and reduce the localization Crámer-Rao bound (CRB) by almost 29\% compared to benchmark algorithms.
\end{abstract}


\section{Introduction}

With the rapid development of unmanned aerial vehicle (UAV) \cite{Lin2021Cooperative,  Chen2024Reconfigurable, b12,  LIN2023249} and artificial intelligence \cite{Lin2024SF, Fu2024Scalable, Lin2024SIC} in the sixth generation (6G) wireless networks, multi-UAVs are increasingly used in military and civil applications. 
Compared with single UAV, cooperative multi-UAVs have a more robust distributed architecture and more comprehensive operational range \cite{b1}. 
Multi-UAV networks can improve the capacity of UAV communications and enhance the reliability of transmissions. 
However, 6G has a higher requirement for security \cite{b15}.
Due to the increasingly serious security issues caused by the ``black flight" of UAVs, incidents harming public property and even national security have occurred many times.

Integrated sensing and communications (ISAC), considered to have a wide range of application prospects, has been established as one of the critical technologies for 6G. In 6G mobile communication networks, ISAC technology will provide high-quality wireless connectivity and high-precision sensing capability, which will boost many emerging applications, such as UAV communication.

Since ISAC can effectively reduce hardware costs and improve spectrum efficiency, researchers have started to explore the application of ISAC in UAV networks. In ISAC-enabled UAV networks, designing efficient joint optimization algorithms for adaptive resource allocation has become a challenge. Liu \emph{et al.} \cite{b3} proposed a periodic pass-sensing integration mechanism, which jointly optimized transmitting beamforming, user association, sensing time selection, and UAV trajectory to maximize the achievable rate. Cao and Yu \cite{b4} investigated a large-scale multiple-input multiple-output ISAC system that supports cell-free and proposed a joint user scheduling and power allocation scheme, where the nonconvex resource allocation problem is relaxed to a linear programming problem. Liu \emph{et al.} \cite{b5} achieved maximum energy efficiency and minimum user radar mutual information for ISAC-enabled UAVs by optimizing user scheduling, transmit power and UAV trajectories. However, the researches mentioned above do not consider multi-UAV cooperative communication and the potential non-cooperative UAV localization problem. In addition, the above works deal with mixed integer nonconvex problems by relaxing the discrete variables into continuous variables, which inevitably causes accuracy problems.

Therefore, avoiding collision or interference to the flight trajectory within UAVs is necessary by localizing the location of non-cooperative UAV. Meanwhile, accurate position acquisition between UAV cluster nodes is the basis of UAV cluster cooperative. 
In addition, 6G also requires the system to have the adaptive capability \cite{b13, Xiao2023Secure }.
How to design efficient joint optimization algorithms for adaptive resource allocation and deal with the mixed integer nonconvex resource allocation problem more efficiently in ISAC-enabled UAV networks has become an important problem. In this paper, we study the resource allocation problem for cooperative communication with non-cooperative localization in the ISAC-enabled multi-UAV cooperative network. We summarize the main contributions in the following:

\begin{itemize}
\item The relationship between localization quality of service (QoS) and sum rate is first revealed. We formulate an optimization problem that maximizes the weighted sum of the system average sum rate and localization QoS by jointly optimizing the optimal cell association, communication power and sensing power allocation under the constraints of the localization QoS of all UAVs and the sum rate of cooperative UAVs.

\item Since the formulated problem is a mixed-integer nonconvex problem and optimal transport theory (OT) can solve mixed-integer nonconvex problems more effectively by utilizing probability theory to capture the generalized distribution, the alternating iteration algorithm based on optimal transport theory (AIBOT) is proposed to solve the optimization problem more effectively.

\item The simulation results demonstrate that the AIBOT can improve the system sum rate by nearly 12\% and reduce the localization Crámer-Rao bound (CRB) by almost 29\% compared to benchmark algorithms.
\end{itemize}

\section{System Model}

\subsection{Network Model}

\begin{figure}
	\centering
	\includegraphics[width=2.7in]{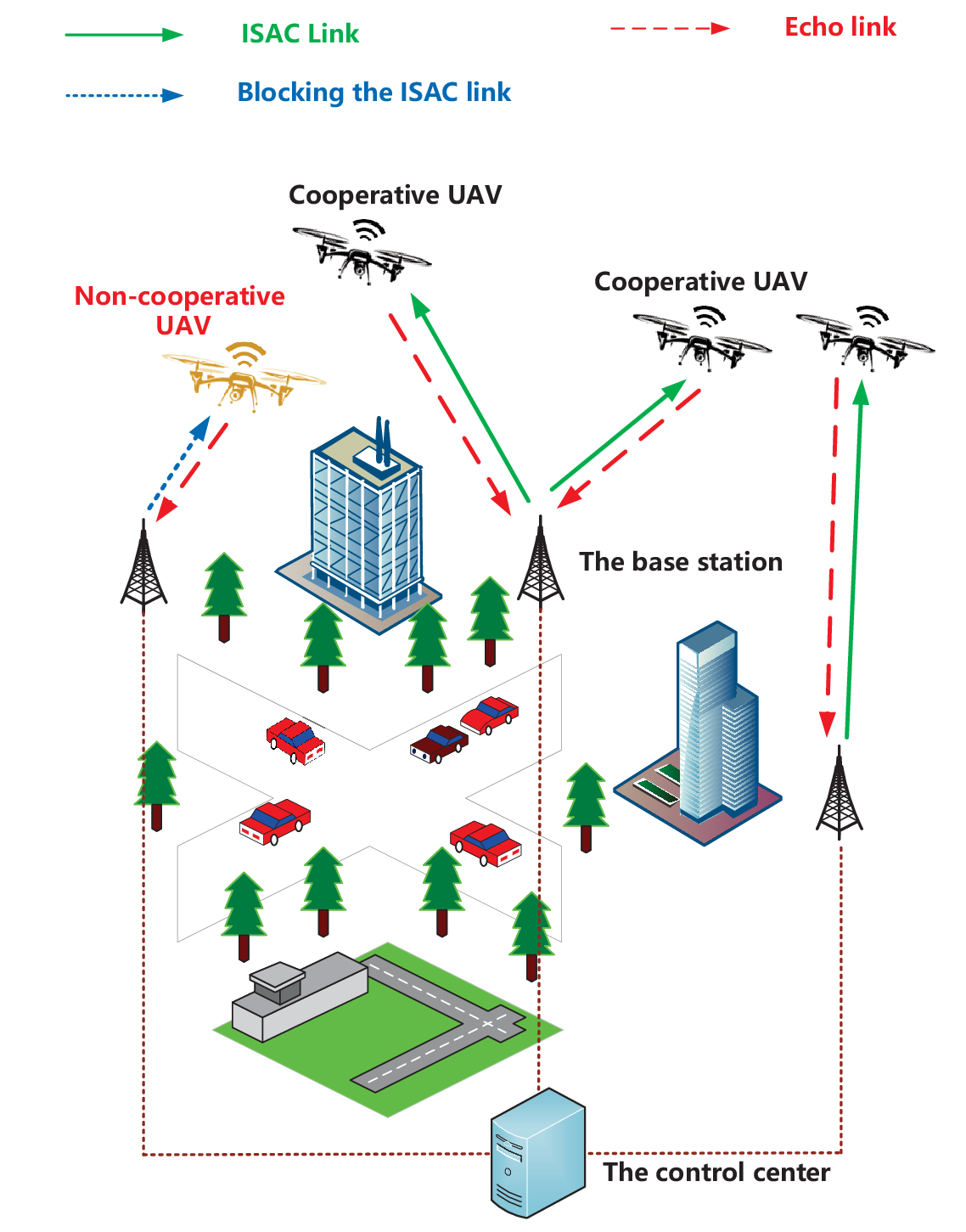}
	\caption{The system model.}
        \label{fig1}
\end{figure}

We consider a downlink communication system supporting ISAC. The system consists of $M$ multi-antenna dual-functional ground base stations, $K + 1$ UAVs, and a control centre, where the UAVs consist of $K$ single-antenna cooperative UAV users and one single-antenna non-cooperative UAV, as shown in Fig. \ref{fig1}. At the $n$-th time slot, the coordinates of the $k$-th UAV are denoted as ${\boldsymbol{\widetilde U}_k}(n) = ({x_{u,k}}(n),{y_{u,k}}(n),{z_{u,k}}(n))$, the coordinates of the  $m$-th dual-function base station are denoted as ${\boldsymbol{\widetilde B}_m}(n) = ({x_{b,k}}(n),{y_{b,k}}(n),{z_{b,k}}(n))$. The coordinates of the non-cooperative UAVs are denoted as ${\boldsymbol{\widetilde A}(n)}$. 

Each base station is equipped with a uniform planar array antenna consisting of ${N_t}$ antenna units, where the position of the $en$-th antenna at the $m$-th base station denotes $\boldsymbol{\pounds}_{m,en}$. Let ${a_{j,m}}(n) \in \{ 0,1\} ,n \in \{ 1, \cdots ,N\}$ denote the association variable between the $m$-th ground base station and the $k$-th UAV. If ${\boldsymbol{\widetilde U}_k} \in {\boldsymbol{Q}_m}$, then ${a_{j,m}}(n) = 1$ indicates that the $k$-th UAV is provided with ISAC or sensing service by the $m$-th base station at the $n$-th time slot. Otherwise, ${a_{j,m}}(n) = 0$.

The total flight time of the UAV is denoted by $T$, which is divided into $N$ time slots having the same time interval ${\delta _t}$.

We define UAVs’ spatial distribution function $f\left( {x,y,z} \right)$ to capture the probability of each UAV appearing around a three-dimensional location $\boldsymbol{q}=\left( {x,y,z} \right)$. ${\boldsymbol{Q}_m}$ denotes the three-dimensional cell associated with the $m$-th dual-functional ground base station, and the average number of UAV inside ${\boldsymbol{Q}_m}$ can be expressed as
\begin{equation}
{U_m} = K\int {_{{\boldsymbol{Q}_m}}f(\boldsymbol{q})d\boldsymbol{q}} ,
\end{equation}
where the spatial distribution function $f(\boldsymbol{q}) = f(x,y,z)$ captures the probability of each UAV user appearing around the three-dimensional location $\boldsymbol{q} = \left( {x,y,z} \right)$.

The departure angle between the $m$-th ground base station and the $k$-th UAV can be denoted as ${\boldsymbol{\theta} _{k,m}} = {[\theta _{k,m}^{az},\theta _{k,m}^{el}]^T}$, where $\theta _{k,m}^{az}$ and $\theta _{k,m}^{el}$ indicate the departure angle in azimuth and elevation angles, respectively. $\theta _{k,m}^{az}$ and $\theta _{k,m}^{el}$ can be expressed by
\begin{align}
\label{eq2}
&\theta _{k,m}^{az} \buildrel \Delta \over = a\tan 2({[{\widetilde B_m}]_2} - {[{\boldsymbol{\widetilde U}_k}]_2},{[{\boldsymbol{\widetilde B}_m}]_2} - {[{\boldsymbol{\widetilde U}_k}]_1}), \nonumber \\
&\theta _{k,m}^{el} \buildrel \Delta \over = \arccos\left (\frac{{{{[{\boldsymbol{\widetilde B}_m}]}_3} - {{[{\boldsymbol{\widetilde U}_k}]}_3}}}{{\| {{\boldsymbol{\widetilde B}_m} - {\boldsymbol{\widetilde U}_k}} \|}}\right).
\end{align}

\subsection{Communication Model}

After completing stages 1 and 2, the base station acquires estimated channel state information ${\boldsymbol{\widehat h}_{k,m}}$ for the UAV and its position ${\boldsymbol{\widehat U}_k}$. ${\boldsymbol{\widehat h}_{k,m}}^c(n)$ by the path loss model can be expressed as
\begin{equation}
\boldsymbol{\widehat h}_{k,m}^c(n) = \frac{{\lambda {e^{j{\varphi _{k,m}}(n)}}}}{{4\pi ({{\widehat d}_{k,m}}(n) + \Delta {d_{k,m}}(n))}}{\boldsymbol{\widehat \alpha} _{k,m}}(n),
\end{equation}
where ${\boldsymbol{\widehat \alpha} _{k,m}}(n)$ represents the estimated response vector of the antenna array, which can be calculated based on the estimated position by (\ref{eq2}). ${\varphi _{k,m}}(n)$ represents the phase shift, $\Delta d_{k,m}(n) = {d_{k,m}}(n) - {\widehat d_{k,m}}(n)$ represents the location sensing error, and ${d_{k,m}}(n) = \left\| {{\boldsymbol{\widetilde B}_m}(n) - {\boldsymbol{\widetilde U}_k}(n)} \right\|$ represents the actual distance between the base station and the UAV.

The communication signal $y_s^{k,m}\left( n \right)$ received by the $k$-th cooperative UAV from the $m$-th base station at the $n$-th time slot using the receiver beamforming vector ${\widetilde w_{k,m}}(n)$ can be expressed as
\begin{align}
y_s^{k,m}\left( n \right) &= \sqrt {p_s^{k,m}} {({\boldsymbol{\widehat h}_{k,m}^c} + \Delta {\boldsymbol{h}_{k,m}^c})^H}{\boldsymbol{w}_{k,m}}s_s^{k,m}(n - {\tau _{k,m}})\nonumber\\
&\quad + b(n),
\end{align}
where $b(n)$ denotes additive white Gaussian noise, $\Delta {\boldsymbol{h}_{k,m}^c}(n)$ denotes the channel state information error, which can be expressed as 
\begin{equation}
\Delta {\boldsymbol{h}_{k,m}^c}(n) = \frac{{\lambda {e^{j{\varphi _{k,m}}}}{{\boldsymbol{\widehat \alpha} }_{k,m}}(n)}}{{4\pi \Delta {d_{k,m}}(n)}} = \frac{{\lambda {e^{j{\varphi _{k,m}}(n)}}{{\boldsymbol{\widehat \alpha} }_{k,m}}(n)}}{{4\pi ({d_{k,m}}(n) - {{\widehat d}_{k,m}}(n))}}.
\end{equation}

The sum rate of all cooperative UAVs can be expressed as 
\begin{equation}
\begin{array}{l}
{R_{sum}} = \sum\limits_{k = 1}^K {\sum\limits_{n = 1}^N {{b_{k,m}}(n)} } \\
\qquad \qquad {\rm{       }}\times {\log _2}[1 + {a_{k,m}}(n)\frac{{p_c^{k,m}(n){\lambda ^2}{{\left| {\boldsymbol{\widehat \alpha} _k^H(n){\boldsymbol{w}_k}(n)} \right|}^2}}}{{16{\pi ^2}{N_0}({{\widehat d}_{k,m}}(n) + \Delta {d_{k,m}}(n))}}],      
\end{array}
\end{equation}
where ${b_{k,m}}(n)$ denotes the bandwidth allocated to the $k$-th UAV by the $m$-th base station in the $n$-th time slot.

\subsection{Sensing Model}

The baseband form $y_s^{j,m}\left( n \right)$ in which the $m$-th ground base station receives the reflected echo signal from the $k$-th UAV can be expressed as
\begin{equation}
y_s^{j,m}\left( n \right) = \sqrt {p_s^{j,m}(n)} \boldsymbol{h}_{j,m}^{{\rm{ s   }}H}(n){\boldsymbol{w}_{j,m}}(n)s_s^{j,m}(n - {\tau _{j,m}}) + b(n),
\end{equation}
where $\boldsymbol{h}_{j,m}^s(n)$ denotes the composite channel between the $k$-th UAV and the $m$-th base station, ${\tau _{j,m}} = \frac{{\left\| {{\boldsymbol{\widetilde B}_m} - {\boldsymbol{\widetilde U}_j}} \right\|}}{c}$ denotes the delay between the $j$-th UAV and the $m$-th base station, and $c$ denotes the speed of light.  $\boldsymbol{h}_{k,m}^s(n)$ can be expressed as
\begin{equation}
\boldsymbol{h}_{j,m}^s(n) = {\varsigma _{j,m}}(n)\boldsymbol{\alpha} ({\boldsymbol{\theta} _{j,m}}(n)),
\end{equation}
where ${\varsigma _{j,m}}(n) = \left| {{\varsigma _{j,m}}(n)} \right|{e^{j{\varphi _{j,m}}(n)}}$ denotes the composite channel gain, $\left| {{\varsigma _{j,m}}(n)} \right| = \frac{\lambda }{{4\pi \left\| {{\boldsymbol{\widetilde U}_j}(n) - {\boldsymbol{\widetilde B}_m}(n)} \right\|}}$, ${\varphi _{j,m}}(n)$ denotes the phase shift, $\boldsymbol{\alpha} ({\boldsymbol{\theta} _{j,m}}(n))$ denotes the antenna response vector, which can be expressed as \cite{b7}
\begin{equation}
\boldsymbol{\alpha} ({\boldsymbol{\theta} _{j,m}}(n)) = {e^{j{\boldsymbol{Q}}_{k,m}^T(n)\boldsymbol{\kappa} ({\theta _{j,m}}(n))}},
\end{equation}
where ${\boldsymbol{Q}_{k,m}}(n) \buildrel \Delta \over = [{\boldsymbol{\pounds}_{m,1}}(n), \cdots ,{\boldsymbol{\pounds}_{m,{N_t}}}(n)]$, and $\boldsymbol{\kappa} ({\boldsymbol{\theta} _{j,m}}(n)) \buildrel \Delta \over = \frac{{2\pi }}{\lambda }[\sin {\theta _{el}}(n) \cos {\theta _{az}}(n),\sin {\theta _{el}}(n)\sin {\theta _{az}}(n), \\ \cos {\theta _{el}}(n)]$, $\lambda$ denotes the wavelength.

A generalized form to simplify the expression is used to characterize the estimated CRB for the $j$-th UAV, which can be expressed as \cite{b8}
\begin{align}
&CRB({d_j}(n)) \propto {(p_s^{j,m}(n){\left| {{\varsigma _j}(n)} \right|^2}B_{rms,j}^2)^{ - 1}}, \nonumber \\
&CRB({\theta _j}(n)) \propto {(p_s^{j,m}(n){\left| {{\varsigma _j}(n)} \right|^2}{W_{NN}})^{ - 1}},
\end{align}
where $p_s^{j,m}(n)$ denotes the transmit power for localization non-cooperative UAV, ${B_{rms,j}}$ and ${W_{NN}}$ represent the effective bandwidth and the null-to-null beam width of the receiving antennas, respectively.

Localization focuses on target distance and angle of arrival estimation, so the localization error is defined as the CRB of target distance and angle of arrival estimation. By transforming the minimization problem of the CRB into the maximization problem of its inverse, the target parameters of $j$-th sensing target can be expressed as
\begin{align}
&{\rm{            }}{\rho _j} = \frac{{{\varpi _1}}}{{CRB(d)}} + \frac{{{\varpi _2}}}{{CRB(\theta )}}  \nonumber \\
&\quad {\rm{               }} = {\varpi _1}\frac{{p_s^{j,m}(n){{\left| {{\varsigma _{j,m}}(n)} \right|}^2}{b_j}}}{{{\beta _1}}} + {\varpi _2}\frac{{p_s^{j,m}(n){{\left| {{\varsigma _{j,m}}(n)} \right|}^2}}}{{{\beta _2}}},
\end{align}
where ${\beta _1}$ and ${\beta _2}$ denote the scaling factor, ${\varpi _1}$ and ${\varpi _2}$ denote the normalization factors used to unify the distance and departure angle units. 

\subsection{Problem Formulation}
For simplicity, time slot $n$ will be omitted in the following. The cell association variable, the communication and sensing power allocation variable can be denoted as $ \boldsymbol{Q} = \{ Q_1^*, \cdots ,Q_M^*\} $, ${ \boldsymbol{P}_c} = \{ p_c^{1,1}, \cdots ,p_c^{K,M},p_c^{K + 1,M}\} $, ${ \boldsymbol{P}_s} = \{ p_s^{1,1}, \cdots ,p_s^{K,M},p_s^{K + 1,m}\} $. The objective is to maximize the weighted sum of the system average sum rate and average localization QoS by jointly optimizing the cell association, communication power allocation and sensing power allocation under non-cooperative UAV localization QoS and cooperative UAV sum rate constraints, which can be formulated as
\begin{align}
\label{eq12}
  &{\textbf{         P1}}:\mathop {\max }\limits_{ \boldsymbol{Q},{ \boldsymbol{P}_c},{ \boldsymbol{P}_s}} {G_{TOL}} = {\vartheta _1}\frac{{{R_{sum}}}}{K} + (1 - {\vartheta _1}){\vartheta _2}\frac{{{\rho _{sum}}}}{{K + 1}} \hfill \nonumber\\
  s.t.&{\text{    C1}}:{U_m} = M\int {_{{\boldsymbol{Q}_m}}f(\boldsymbol{q})d\boldsymbol{q}} , \hfill \nonumber\\
  &{\text{         C2}}:{\boldsymbol{Q}_r} \cap {\boldsymbol{Q}_i} = \emptyset ,\forall i \ne r \in \{ 1, \cdots ,M\} ,\mathop  \bigcup \limits_{m \in M} {\boldsymbol{Q}_m} = \boldsymbol{q}, \hfill \nonumber \\
  &{\text{         C3}}:{p_{\min }} \leqslant p_c^{k,m},p_s^{k,m} \leqslant {p_{\max }}, \forall k \in \{ 1, \cdots ,K\} ,\nonumber \\
  &\qquad {\text{                }}\forall m \in \{ 1, \cdots ,M\} , \hfill \nonumber\\
  &{\text{         C4}}:\sum\limits_{k = 1}^K {{p_{k,m}}}  = {P_{total}},\forall m \in \{ 1, \cdots ,M\} , \hfill \nonumber\\
  &{\text{         C5}}:{\rho _k} \leqslant {\rho _{\min }}{\text{,}}\forall k \in \{ 1, \cdots ,K\} \hfill \nonumber\\
  &{\text{         C6}}:{R_{\min }}{\text{ }} \leqslant {R_k}{\text{,}}\forall k \in \{ 1, \cdots ,K\} 
\end{align}
where ${R_{sum}}$ and ${\rho _{sum}}$ represent the sum rate of all cooperative UAVs and target parameters of all UAVs, respectively. ${\vartheta _1}$ denotes the weights of sum rate and localization QoS and ${\vartheta _2}$ denotes the sum rate-localization QoS mapping factor to ensure that the sum rate and localization QoS are at the same scale level \cite{b8}. C1 denotes the number of cooperative UAV users in the region ${Q_m}$. C2 ensures that the three-dimensional correlation spaces are disjoint and their combination covers the entire space under consideration. C3 denotes the maximum and minimum transmission communication power and sensing power constraints. C4 denotes the total power budget constraints. C5 denotes the UAV localization QoS constraints. C6 denotes the communication QoS constraints.

\section{The AIBOT-Based Joint Cell Association and Power Allocation Optimization}
\textbf{P1} is a mixed integer nonconvex problem and the objective function as well as constraints are highly nonconvex, which is difficult to convert into a convex function. However, OT can utilize probability theory and statistics to capture the generalized distribution of wireless devices. It can provide a more in-depth fundamental analysis of network performance. Therefore, we propose AIBOT to solve the proposed problem.

Since the OT is used to solve the minimization problem, \textbf{P1} can be transformed into
\begin{align}
&{\textbf{P2}}:\mathop {\min }\limits_{ \boldsymbol{Q},{ \boldsymbol{P}_c},{ \boldsymbol{P}_s}} {\rm{ }} - {{\rm{G}}_{TOL}} \nonumber\\
{\rm{       }}s.t. &\quad C1,C2,C3,C4,C5,C6.
\end{align}

\subsection{Cell Association Design} 
For a given transmit power, let $z({U_l}) = \frac{1}{{{U_l}}}$, and the optimal cell association optimization subproblem can be expressed as
\begin{align}
&\textbf{P3}: \mathop{\min}\limits_{\boldsymbol{Q}} \sum_{m=1}^M \int_{\boldsymbol{Q}_m} \frac{1}{{U_m}} \left[ G_{\text{TOL}} f(x,y,z) \right] dxdydz \nonumber \\
&\qquad \qquad{\rm{                = }}\sum\limits_{m = 1}^M {\int {_{{\boldsymbol{Q}_m}}z({U_m})[{G_{TOL}}f(x,y,z)]dxdydz} } \nonumber \\
&\qquad s.t. \quad C1,C2,C5,C6.
\end{align}

\textbf{P3} can be modeled as a semi-discrete OT problem, where the base stations are considered to follow discrete distributions, and the UAVs are considered to follow continuous distributions. The optimal cell association must be obtained by mapping UAVs to the base stations at a minimum cost. Next, we can proceed to solve \textbf{P3} using the following theorem:

\begin{theorem}  The optimal 3D cellular association for a dual-functional ground base station can be expressed as
\begin{equation}
{\boldsymbol{q}_M} = \ \left\{ \mathop {\arg \min }\limits_{i \in \{ 1, \cdots ,M\} }  - \frac{1}{{{U_i}}}{G_{TOL}}({\boldsymbol{q}_0}) + \frac{{{G_{T{\rm{OL}}}}({\boldsymbol{q}_0})}}{{{U_i}K}}\ \right\} .
\label{eq15}
\end{equation}
\end{theorem} 
\vspace{-1\baselineskip}
\begin{proof} The existence of an optimal cell association ${\boldsymbol{Q}_m}$ can be verified by the existence of an OT mapping \cite{b12}. There exist two cells ${\boldsymbol{Q}_i}$ and ${\boldsymbol{Q}_r}$ and a point ${\boldsymbol{q}_0} = ({x_0},{y_0},{z_0}) \in {\boldsymbol{Q}_i}$. Let ${\boldsymbol{O}_\chi }({\boldsymbol{q}_0})$ be a ball centered at ${\boldsymbol{q}_0}$ with radius $\chi  > 0$. Generate the following new three-dimensional partition ${\boldsymbol{\widetilde Q}_m}$, which can be denoted as
\begin{align}
&{\boldsymbol{\widetilde Q}_i} = {\boldsymbol{Q}_l}/{\boldsymbol{O}_\chi }({\boldsymbol{q}_0}),\nonumber \\
&{\boldsymbol{\widetilde Q}_r} = {\boldsymbol{Q}_r} \cup {\boldsymbol{O}_\chi }({\boldsymbol{q}_0}),\nonumber \\
&{\boldsymbol{\widetilde Q}_m} = {\boldsymbol{Q}_m},m \ne i,r.
\end{align}

The set ${\boldsymbol{Q}_m}$ is the optimal cellular association. Therefore, the optimal cell association obtained from ${ \boldsymbol{\widetilde Q}_m}$ will not be better than that obtained from ${\boldsymbol{Q}_m}$. It can be obtained as
\begin{align}
&\quad \int {_{{\boldsymbol{Q}_i}}[z({U_i}) - z({U_i} - {U_\chi })]{G_{TOL}}f(\boldsymbol{q})d\boldsymbol{q}} \nonumber \\
&\quad + \int {_{{\boldsymbol{O}_\chi }({\boldsymbol{q}_0})}z({U_i} - {U_\chi }){G_{TOL}}f(\boldsymbol{q})d\boldsymbol{q}} \nonumber \\
&\le  \int {_{{\boldsymbol{Q}_r}}[z({U_r} + {U_\chi }) - z({U_r})]{G_{TOL}}f(\boldsymbol{q})d\boldsymbol{q}} \nonumber \\
&\quad + \int {_{{\boldsymbol{O}_\chi }({\boldsymbol{q}_0})}z({U_r} + {U_\chi }){G_{TOL}}f(\boldsymbol{q})d\boldsymbol{q}} 
\label{eq17}
\end{align}
\vspace{-1\baselineskip}

Multiplying $\frac{1}{{{U_\chi }}}$ on both sides of (\ref{eq17}). Then, we take the limit when $\chi  \to 0$, and it can be calculated to obtain ${U_\chi } = K\int {_{{\boldsymbol{Q}_\chi }}f(\boldsymbol{q})d\boldsymbol{q}} $, $z'({U_i}) = \frac{{dz(t)}}{{dt}}\left| {^{t = {U_i}}} \right. =  - \frac{1}{{U_i^2}}$, $z'({U_r}) = \frac{{dz(t)}}{{dt}}\left| {^{t = {U_r}}} \right. =  - \frac{1}{{U_r^2}}$. We can get
	\begin{align}
		&- \frac{1}{{U_i^{}}}{G_{TOL}}({\boldsymbol{q}_0}) + \frac{{{G_{TOL}}(x,y,z)}}{{{U_i}K}}\nonumber \\
		\le  & - \frac{1}{{U_r^{}}}{G_{TOL}}({\boldsymbol{q}_0}) + \frac{{{G_{TOL}}(x,y,z)}}{{{U_r}K}}.
	\end{align}

The optimal cell association for the $m$-th ground base station can be expressed as
\begin{equation}
\begin{array}{l}
{\boldsymbol{Q}}_m^* = \ \left\{  (x,y,z)\left| { - \frac{1}{{{U_i}}}{G_{TOL}}(x,y,z) + \frac{{{G_{T{\rm{OL}}}}(x,y,z)}}{{{U_i}K}}} \right. \right.\\
\qquad\qquad \left. \le  - \frac{1}{{{U_r}}}{G_{TOL}}(x,y,z) + \frac{{{G_{TOL}}(x,y,z)}}{{{U_r}K}},\forall i \ne m\ \right\} .	
\end{array}
\end{equation}
which completes the proof of Theorem 1. \end{proof}

Based on the conclusion of Theorem 1, an iterative algorithm based on \cite{b11} is proposed, as shown in Algorithm \ref{algorithm1}, where $\Re _m^{^{({t_1})}}(\boldsymbol{q})$ denotes the predefined parameters used for cell partitioning.

\vspace{-0.5\baselineskip}
\begin{algorithm} 
	\caption{The iterative algorithm for searching for optimal cell associations}
	\begin{algorithmic}[1]
 \label{algorithm1}
		\REQUIRE  $f(x,y,z)$, ${T_1}$, $K$, ${\boldsymbol{B}_m}(n)$
		\ENSURE  $\boldsymbol{Q}_m^*,m \in M$
		
		\textbf{Initialization:}	${t_1} = 1$, generate initial cell associations $Q_m^{({t_1})}$, $\Re _m^{^{({t_1})}}(\boldsymbol{q}) = 0, \boldsymbol{q}=(x,y,z),\forall m \in \{ 1, \cdots ,M\}$
		\WHILE{${t_1} < {T_1}$}
	\STATE Computing \newline $\Re _m^{({t_1} + 1)}(\boldsymbol{q}) \!=\! \left\{ \begin{array}{l}
\!\!\!\![1 - 1/{t_1}]\Re _m^{({t_1})}({\boldsymbol{q}}){\rm{             }},if{\rm{ }}(\boldsymbol{q}) \in \boldsymbol{Q}_m^{({t_1})}\\
\!\!\!\! 1 - [1 - 1/{t_1}][1 - \Re _m^{({t_1})}(\boldsymbol{q})],otherwise
\end{array} \right.$
        \STATE Computing ${U_m} = \int {_\textbf{Q}} (1 - \Re _m^{({t_1})}(\boldsymbol{q}))f(x,y,z)dxdydz$
        \STATE $t \to t + 1$
        \STATE Updating cell associations by (\ref{eq15})
        \ENDWHILE
	\end{algorithmic}
\end{algorithm}
\vspace{-0.6\baselineskip}

\subsection{Transmitting Power Design}

For the given cell association, the transmit power optimization subproblem can be expressed as
\begin{align}
&{\textbf{P4}}:\mathop {\min }\limits_{{ \boldsymbol{P}_c},{ \boldsymbol{P}_s}} {\rm{ }} - {{\rm{G}}_{TOL}}\nonumber \\
{\rm{       }}s.t.&\quad C3,C4,C5,C6.
\end{align}

\textbf{P4} involves finding the OT scheme with optimal transmit power. The source distribution is the transmit power allocated by the base station, the destination distribution is the power received by the UAV and the transmission cost is the negative of the objective function \cite{b9}.

According to the Monge-Kantorovich problem \cite{b8}, \textbf{P4} can be rewritten as
\begin{align}
& \mathop {\max }\limits_{\varphi ,\phi } \int {\psi ({ \boldsymbol{P}_c}){\rm{ }}} \Omega ({ \boldsymbol{P}_c})d{ \boldsymbol{P}_c} + \int {\phi ({ \boldsymbol{P}_s}){\rm{ }}} ({ \boldsymbol{P}_s})d{ \boldsymbol{P}_s} \nonumber\\
& = \mathop {\max }\limits_{\varphi ,\phi } \int {\frac{1}{K}\sum\limits_{k = 1}^K {{{\log }_2}} \left( {1\hspace{-0.1cm} +\hspace{-0.08cm} {a_{k,m}}\frac{{p_c^{k,m}{\lambda ^2}{{\left| {\boldsymbol{\widehat \alpha} _k^H{\boldsymbol{w}_k}} \right|}^2}}}{{16{\pi ^2}{N_0}({{\widehat d}_{k,m}} + \Delta {d_{k,m}})}}} \right)}\nonumber\\ 
&\qquad \qquad \qquad  \cdot{\vartheta _1}{b_{k,m}} d{ \boldsymbol{P}_c}\nonumber \\
&\qquad + \int {\rm{ }} \frac{1}{{K + 1}}\sum\limits_{j = 1}^{K + 1} {\left[ {\left( {{\varpi _1}\frac{{{b_j}}}{{{\beta _1}}} + {\varpi _2}\frac{1}{{{\beta _2}}}} \right)\left( {p_s^{j,m}{{\left| {{\varsigma _j}} \right|}^2}} \right)} \right]} d{ \boldsymbol{P}_s}\nonumber \\
&s.t.  {\rm{  }}\psi ({ \boldsymbol{P}_c})\hspace{-0.1cm} =\hspace{-0.1cm} \frac{1}{K}\sum\limits_{k = 1}^K {{{\log }_2}} \hspace{-0.1cm} \left( {1 \hspace{-0.1cm}+\hspace{-0.08cm} {a_{k,m}}\frac{{p_c^{k,m}{\lambda ^2}{{\left| {\boldsymbol{\widehat \alpha} _k^H{\boldsymbol{w}_k}} \right|}^2}}}{{16{\pi ^2}{N_0}({{\widehat d}_{k,m}} + \Delta {d_{k,m}})}}} \right),\nonumber \\
&{\rm{       }}\Omega ({ \boldsymbol{P}_c}) = \frac{1}{K}\sum\limits_{k = 1}^K {{\vartheta _1}{b_{k,m}}} , \nonumber \\
&{\rm{       }}\phi ({ \boldsymbol{P}_s}) = \frac{1}{{K + 1}}\sum\limits_{j = 1}^{K + 1} {\left( {p_s^{j,m}{{\left| {{\varsigma _j}} \right|}^2}} \right)} ,\nonumber \\
&{\rm{       }}\mho ({ \boldsymbol{P}_s}) = \frac{1}{{K + 1}}\sum\limits_{j = 1}^{K + 1} {\left( {{\varpi _1}\frac{{{b_j}}}{{{\beta _1}}} + {\varpi _2}\frac{1}{{{\beta _2}}}} \right)} .
\label{eq21}
\end{align}

To maximize ${G_{TOL}}$, $\psi $ and $\phi $ must be maximized within the expected range. Since it is necessary for any ${ \boldsymbol{P}_c} \in {D_1}$ and ${ \boldsymbol{P}_s} \in {V_1}$ to satisfy $\psi ({ \boldsymbol{P}_c}) + \phi ({ \boldsymbol{P}_s}) \le  - {\vartheta _1}\frac{1}{K}{R_{sum}} - (1 - {\vartheta _1})\frac{1}{{K + 1}}{\vartheta _2}{\rho _{sum}}$, after determining $\psi $, $\phi $ can be represented as
\begin{equation}
\phi ({ \boldsymbol{P}_s}) = \psi ({ \boldsymbol{P}_c}) = \inf \{  - {G_{TOL}}({ \boldsymbol{P}_c},{ \boldsymbol{P}_s})\}  - \psi ({ \boldsymbol{P}_c}),
\end{equation}
where $\inf $ represents the infimum of the function.

$\phi ({ \boldsymbol{P}_s})$ can be simplified as
\begin{equation}
\mathop {\max }\limits_{\psi ,\phi } F(\phi ) = \mathop {\max }\limits_{\varphi ,\phi } \int {\psi ({ \boldsymbol{P}_c}){\rm{ }}} \Omega ({ \boldsymbol{P}_c})d{ \boldsymbol{P}_c} + \int {\phi ({ \boldsymbol{P}_s}){\rm{ }}} ({ \boldsymbol{P}_s})d \boldsymbol{P}_s,
\label{eq23}
\end{equation}
where $F\left( \phi  \right)$ is the function name. Solving the optimization problem of (\ref{eq23}) will result in a set of $\phi$, by which the optimal power allocation scheme  $ \boldsymbol{P}_s^i$ can be obtained. Calculating the first-order derivative can solve the optimization problem (\ref{eq23}) using the gradient descent method. It can be expressed as
\begin{align}
\nabla F(\phi ) &= \frac{{\partial F(\phi )}}{{\partial {\phi _i}}} \nonumber\\
&{\rm{            =  }}\int {\Omega ({ \boldsymbol{P}_c})} d{ \boldsymbol{P}_c} + \int {({ \boldsymbol{P}_s})} d{ \boldsymbol{P}_s}\nonumber \\
&{\rm{            =  }}\frac{1}{K}\sum\limits_{k = 1}^K {{\vartheta _1}{b_{k,m}}} { p_c^{k,m}} \nonumber \\
&\quad +\frac{1}{{K + 1}}\sum\limits_{j = 1}^{K + 1} {\left( {{\varpi _1}\frac{{{b_j}}}{{{\beta _1}}} + {\varpi _2}\frac{1}{{{\beta _2}}}} \right)} { p_s^{j,m}}.
\end{align}

Therefore, the power allocation algorithm based on OT is shown in \textbf{Algorithm \ref{algorithm2}}.

\begin{algorithm}
	\caption{The power allocation algorithm based on OT}
	\label{algorithm2}
	\begin{algorithmic}[1]
		\REQUIRE  ${\vartheta _1}$, ${\vartheta _2}$, $\psi ({ \boldsymbol{P}_c}),\phi ({ \boldsymbol{P}_s}) \in D$, $\Omega ({ \boldsymbol{P}_c}),({ \boldsymbol{P}_s}) \in V$
		\ENSURE  $ \boldsymbol{P}_c^* =  \boldsymbol{P}_c^{{t_2}}$, $ \boldsymbol{P}_s^* =  \boldsymbol{P}_s^{{t_2}}$
		
		\textbf{Initialization:}	${t_2} = 1$, ${\phi _t}$
		\WHILE{${\left\| {\nabla F({\phi _t})} \right\|_2} \ge {\vartheta _2}$}
		\STATE $n = 1$, ${m_1} = 1$
		\STATE Updating ${\phi _{t + 1}} = {\phi _t} + m\nabla F({\phi _t})$
		\IF {$\nabla F({\phi _t}) < \nabla F({\phi _{t + 1}})$} 
		\WHILE{$\nabla F({\phi _t}) < \nabla F({\phi _{t + 1}})$}
		\STATE$n \to n + 1$, ${m_n} = {2^{n - 1}}{m_1}$
		\STATE Updating ${\phi _{t + 1}} = {\phi _t} + {m_n}\nabla F({\phi _t})$
		\ENDWHILE
		\ELSE 
		\WHILE{$\nabla F({\phi _t}) > \nabla F({\phi _{t + 1}})$}
		\STATE$n \to n + 1$, ${m_n} = {2^{n - 1}}{m_1}$
		\STATE Updating ${\phi _{t + 1}} = {\phi _t} + {m_n}\nabla F({\phi _t})$
		\ENDWHILE
		\ENDIF  
		\STATE $t \to t + 1$
		\ENDWHILE
	\end{algorithmic}
\end{algorithm}

\subsection{The alternating iteration algorithm based on OT}
AIBOT alternately solves \textbf{P3} and \textbf{P4} by \textbf{Algorithm \ref{algorithm1}} and \textbf{Algorithm \ref{algorithm2}} to solve \textbf{P1}, respectively, as shown in \textbf{Algorithm \ref{algorithm3}}.

\begin{algorithm}
	\caption{The alternating iteration algorithm based on OT}
        \label{algorithm3}
	\begin{algorithmic}[1]
		\REQUIRE  ${T_3}$
		\ENSURE ${ \boldsymbol{Q}^*} = { \boldsymbol{Q}^{t3}}$, $ \boldsymbol{P}_c^* =  \boldsymbol{P}_c^{{t_3}}$, $ \boldsymbol{P}_s^* =  \boldsymbol{P}_s^{{t_3}}$
		
		\textbf{Initialization:}	${t_3} = 1$
		\WHILE{
${t_3} < {T_3}$}
	\STATE $n = 1$, ${m_1} = 1$
        \STATE Given the transmit power, solve \textbf{P3} by \textbf{Algorithm \ref{algorithm1}}
       \STATE Given cell association, solve \textbf{P4} by \textbf{Algorithm \ref{algorithm2}} 
        \ENDWHILE
	\end{algorithmic}
\end{algorithm}

\section{Numerical Results}
This section analyzes the effectiveness of AIBOT in ISAC-enabled multi-UAV cooperative scenarios through simulation. 

\begin{table}[!tb]
\caption{{Parameters Setting}}
\label{table1}
\begin{center}
\renewcommand{\arraystretch}{1.3}{
\begin{tabular}{|c|c|c|c|}
\hline
\textbf{Parameter} & \textbf{Value} & \textbf{Parameter} & \textbf{Value}\\
\hline
\ ${f_c}$ & 30 GHz    &  ${N_0}$ & $-$169 dBm/Hz  \\
\hline
\ ${p_{\min }},{p_{\max }}$ & 0, 40 dBm   &  ${\beta _0}$ &  $-$50 dBW   \\
\hline
\ $M$ & $6$   &  $K$   & $18$  \\
\hline
\end{tabular}}
\label{tab1}
\end{center}
\end{table}

The parameter settings are shown in Table \ref{table1} \cite{b3,b4,b5}. To verify the effectiveness of the alternating iteration algorithm based on the OT, we use the Weighted Voronoi diagram \cite{b12} and the Iterative Water-filling  algorithm as comparison algorithms.

\begin{figure}
	\centering
	\includegraphics[width=2.7in]{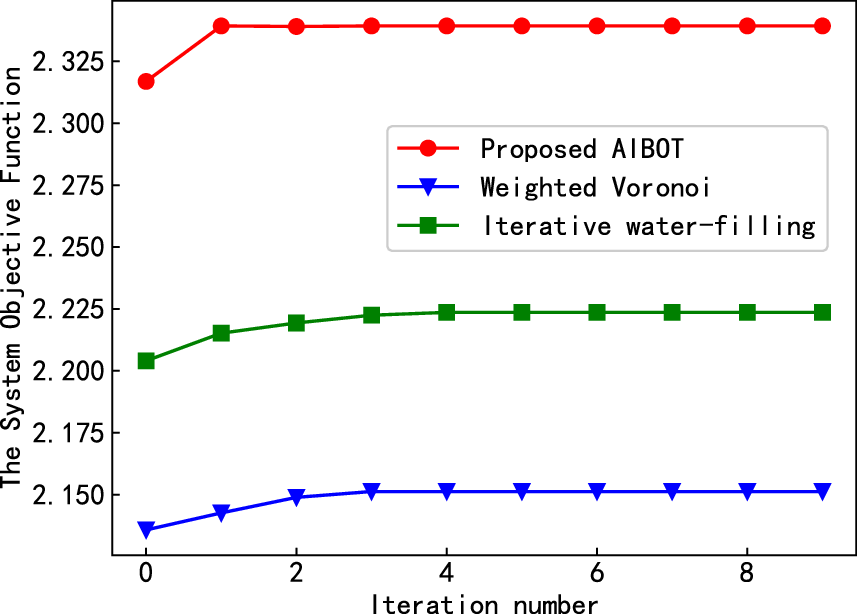}
	\caption{The system objective function versus different number of iterations.}
    \label{Fig2}
\end{figure}

The relationship between the system objective function and the number of iterations is shown in Fig. \ref{Fig2}. As iterations increase, the objective function rises and stabilizes after the 3rd iteration. Compared to the Weighted Voronoi diagram and Iterative Water-filling algorithm, AIBOT improves the objective function by nearly 0.116. This improvement is due to solving the mixed-integer nonconvex problem without relaxing discrete variables, thus avoiding accuracy errors. Additionally, OT's advantage lies in handling mixed integer planning problems by mapping continuous probability density functions to discrete measures in semi-discrete OT problems.

\begin{figure}
	\centering
	\includegraphics[width=2.7in]{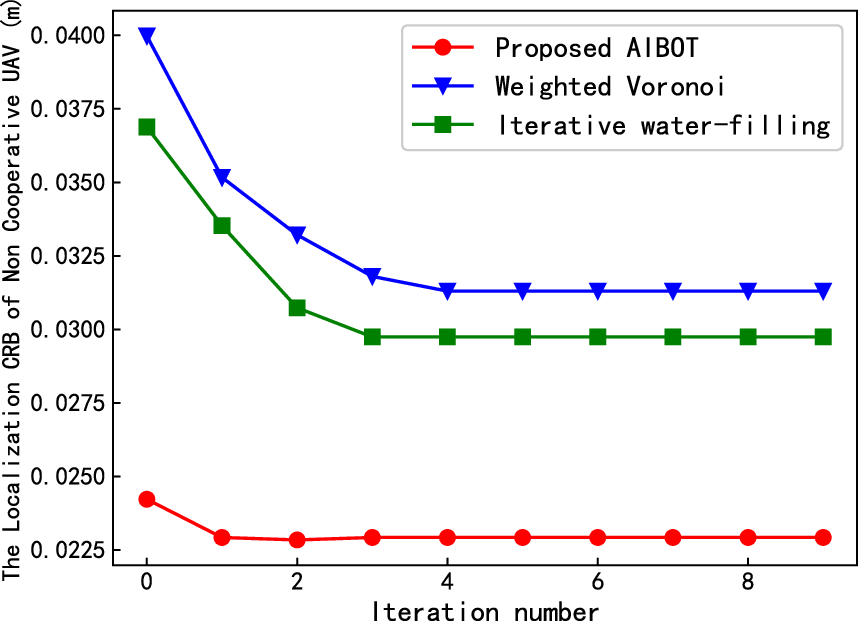}
	\caption{The CRB for non-cooperative UAV localization versus different number of iterations.}\label{Fig3}
\end{figure}

The relationship between the non-cooperative UAV localization CRB and the number of iterations is given in Fig. \ref{Fig3}. It can be observed that the non-cooperative UAV localization CRB gradually decreases with the increase of the number of iterations and gradually stabilizes after the 3rd iteration. It can be demonstrated that the AIBOT can improve the localization CRB by almost 29\%. Therefore, the proposed algorithm can achieve high localization QoS for the non-cooperative UAV.

\begin{figure}
\vspace{-0.8em}
	\centering
	\includegraphics[width=2.7in]{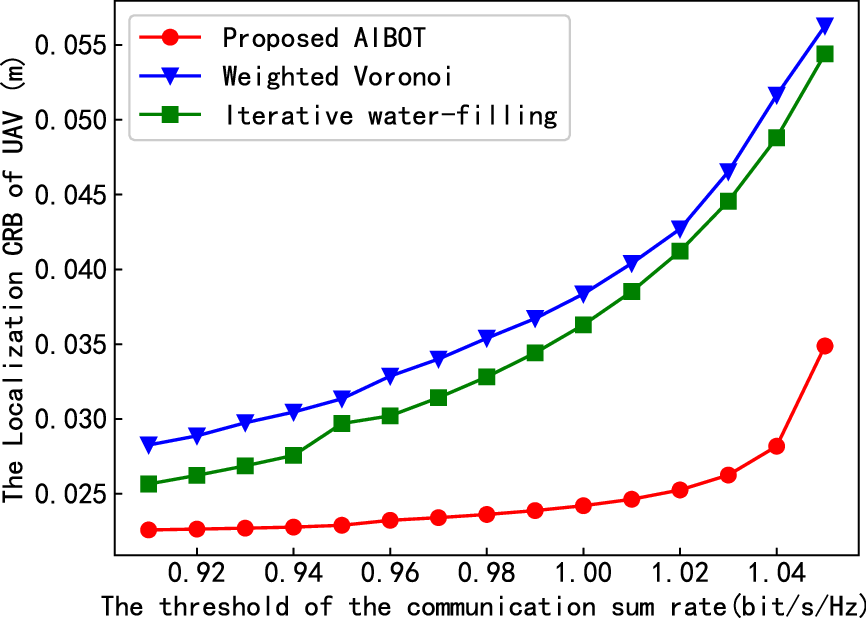}
	\caption{The localization CRB of UAV versus the threshold of the communication sum rate.}\label{Fig4}
\end{figure}

The average sum rate comparison of different algorithms converging in the ISAC-enabled multi-UAV network with different communication rate threshold requirements is illustrated in Fig. \ref{Fig4}. It can be seen that the average sum rate of the system continues to decrease as the communication rate threshold increases. This trend occurs because the total power and channel resources of the ground dual-functional base station are finite. When the demand for sensing increases, the ground dual-functional base station must allocate some of its resources to meet the sensing requirements, consequently reducing the resources available for communication and leading to an increase in the localization CRB of the UAV.

\section{Conclusions}

This paper studies the resource allocation problem for cooperative communication with non-cooperative localization in ISAC-enabled multi-UAV cooperative networks. The relationship between localization QoS and sum rate is revealed. The objective of maximizing the weighted sum of the system average sum rate and localization QoS is achieved by jointly optimizing the cell association, communication power allocation and sensing power allocation. We propose the AIBOT to solve the mixed-integer nonconvex problem. Simulation results demonstrate that the proposed AIBOT can achieve high-rate communication and high-accuracy localization in ISAC-enabled multi-UAV networks.

\bibliographystyle{IEEEtran}
\bibliography{library_abbreviated}

\end{document}